\documentclass[twoside,11pt]{article}

\usepackage{amsmath,amssymb,amsfonts}
\usepackage{graphicx}
\usepackage{graphics}
\usepackage{hyperref}

\begin{document}
\setlength{\unitlength}{25mm}

% ---------------- Title ----------------
\title{Kapitza Dynamics as a New Stabilization Mechanism for Heavy Tetraquarks}

\author{
    M.~Monemzadeh\thanks{monem@kashanu.ac.ir},
    \large N.~Tazimi\thanks{tazimi@kashanu.ac.ir}
    \\\\
    \it\small{Department of Physics, University of Kashan, Kashan, Iran}
}

\maketitle

% ---------------- Abstract ----------------
\begin{abstract}
We investigate a Kapitza-inspired mechanism in which rapid oscillations in the heavy-quark
interaction generate an effective short-range repulsive term in the diquark--antidiquark
potential. The resulting $1/r^{4}$ contribution prevents collapse at short distances and
produces a stable minimum in the effective potential. Within a diquark--antidiquark picture,
we construct a modified Cornell-type potential and analyze the spectrum of heavy tetraquarks
using a Gaussian variational method. We compute the binding energies, wave functions, radii,
and mass spectra of charm and bottom tetraquarks, including the $X(3872)$, $T_{bb}$, and fully
heavy $bb\bar{b}\bar{b}$ states. The model reproduces the mass of the $X(3872)$ and predicts a deeply bound
$T_{bb}$ state consistent with lattice QCD. The fully heavy $bb\bar{b}\bar{b}$ mass also agrees with recent
lattice determinations. Our results indicate that the Kapitza mechanism provides a natural and
robust stabilization effect in multiquark systems and offers a unified description of
molecular-like and compact tetraquark configurations.
\end{abstract}

% ---------------- Introduction ----------------
\section{Introduction}

The discovery of exotic hadrons over the past two decades has profoundly changed our
understanding of the QCD spectrum. The observation of the $X(3872)$ by the Belle
Collaboration and its subsequent confirmation by other experiments opened the door to a rich
family of states that do not fit easily into the conventional quark--antiquark or three-quark
classification schemes \cite{Choi2003,Guo2018}. Additional candidates, such as the $Z_c(3900)$,
$Z_b(10610)$, $T_{cc}$, and several fully heavy states, have further highlighted the importance
of multiquark dynamics and the need for reliable theoretical frameworks. A comprehensive
review of heavy-quark exotic states can be found in Ref.~\cite{Lebed2017}.
Among the various approaches proposed to describe these states, the diquark--antidiquark
picture has emerged as a particularly appealing one. In this framework, two quarks form a
compact diquark in a color $\bar{\mathbf{3}}$ configuration, which interacts with an antidiquark
in a color $\mathbf{3}$ state. This picture provides a natural way to organize the quantum
numbers of tetraquarks and has been used extensively in phenomenological studies
\cite{Brambilla2005,GodfreyIsgur1985,Richard2022}. However, when conventional Cornell-type
potentials are applied directly to diquark--antidiquark systems, one often encounters a serious
difficulty: the short-distance behavior of the potential leads to collapse of the wave function
and the absence of a stable minimum. In other words, the attractive Coulomb term dominates at
small $r$, and the system does not support a well-defined bound state.

This problem suggests that additional short-range physics, beyond the standard
Coulomb-plus-linear interaction, may play an essential role in stabilizing multiquark
configurations. In this work, we explore a mechanism inspired by Kapitza's classical analysis
of rapidly oscillating systems \cite{Kapitza1951,LandauLifshitz1960}. In Kapitza's original
problem, a pendulum with a rapidly oscillating pivot acquires an effective stabilizing potential
that can support an inverted equilibrium. The key idea is that high-frequency oscillations,
when averaged over time, generate an additional effective term in the potential that can be
either stabilizing or destabilizing, depending on the details of the motion.

We adapt this idea to the heavy-quark sector by introducing a rapidly oscillating component
in the interaction between the diquark and antidiquark. Upon averaging over the fast
oscillations, an effective short-range repulsive term emerges, which behaves as $1/r^{4}$ at
small distances. This term provides a strong repulsive core that prevents collapse and allows
the potential to develop a stable minimum. The resulting effective potential can then support
bound tetraquark states in a controlled and physically motivated way.

The main goals of this paper are: (i) to construct a Kapitza-modified potential for
diquark--antidiquark systems, (ii) to derive the corresponding effective $1/r^{4}$ term and
clarify its physical origin, (iii) to analyze the spectrum of heavy tetraquarks using a Gaussian
variational method, and (iv) to compare the resulting mass predictions with experimental and
lattice QCD results for states such as $X(3872)$, $T_{bb}$ and $bb\bar{b}\bar{b}$. It is important to emphasize that,
unlike meson–antimeson molecular approaches, the present work adopts a diquark–antidiquark
picture throughout.

In recent years, several studies have explored multiquark systems using a variety of
nonperturbative techniques, including meson–antimeson molecular models,
Lippmann–Schwinger and Bethe–Salpeter formalisms, relativistic diquark–antidiquark
approaches, and analytical solutions of the Schrödinger equation for both tetraquark
and hexaquark configurations. Representative examples include the meson–antimeson
spectroscopy based on one-pion exchange~\cite{TazimiEPJC2026}, the analytical and
numerical investigations of hexaquark bound states~\cite{TazimiPramana2026,TazimiHexaBS},
the relativistic treatment of heavy tetraquarks~\cite{TazimiEPJC2025},
the Regge-trajectory analysis of multiquark states~\cite{TazimiRegge2025}, and the
Lippmann–Schwinger study of diquark–antidiquark bound systems~\cite{TazimiPLB2015}.
These works collectively demonstrate the effectiveness of two-body reductions and
variational or integral-equation methods in describing multiquark dynamics. The present
study extends this line of research by introducing a Kapitza-inspired short-range
stabilization mechanism for heavy tetraquarks.

\section{Theoretical Model}

We consider a diquark--antidiquark system in the relative coordinate $r$ with reduced mass
$\mu$. The effective Hamiltonian is taken to be
\begin{equation}
H = -\frac{1}{2\mu}\nabla^{2} + V_{\rm eff}(r),
\label{eq:H_general}
\end{equation}
where $V_{\rm eff}(r)$ is an effective potential that incorporates both the standard heavy-quark
interaction and the Kapitza-induced correction.

In the absence of rapid oscillations, the interaction between the diquark and antidiquark is
modeled by a Cornell-type potential,
\begin{equation}
V_{0}(r) = -\frac{a}{r} + \sigma r,
\label{eq:Cornell}
\end{equation}
where $a$ is an effective Coulomb coefficient and $\sigma$ is the string tension. This form has
been widely used in quarkonium and heavy hadron spectroscopy \cite{Eichten1978,Brambilla2005}.
The Coulomb term originates from one-gluon exchange, while the linear term represents the
confining interaction.

However, when applied directly to diquark--antidiquark systems, the potential in
Eq.~\eqref{eq:Cornell} can lead to an unphysical collapse at short distances. The attractive
$-a/r$ term dominates as $r\to 0$, and the wave function tends to concentrate at the origin,
driving the energy to $-\infty$ in the absence of additional short-range physics. This behavior
indicates that the effective interaction in multiquark systems must contain a repulsive
component at very short distances.

To address this issue, we introduce a Kapitza-inspired correction to the potential. The idea
is to allow for a rapidly oscillating component in the interaction, which, upon averaging over
time, generates an additional short-range term. The resulting effective potential takes the form
\begin{equation}
V_{\rm eff}(r) = -\frac{a}{r} + \sigma r + V_{K}(r),
\label{eq:Veff_withVK}
\end{equation}
where $V_{K}(r)$ is the Kapitza-induced contribution. In the next section we derive the explicit
form of $V_{K}(r)$ and show that it behaves as $1/r^{4}$ at short distances.

The parameters $a$ and $\sigma$ are chosen in line with standard heavy quarkonium
phenomenology. In the numerical analysis we use
\begin{equation}
a = 0.50, \qquad \sigma = 0.18~\mathrm{GeV}^{2},
\end{equation}
which are typical values in potential models for heavy quark systems
\cite{GodfreyIsgur1985,Brambilla2005}. The Kapitza coefficient $K$ appearing in $V_{K}(r)$
will be fixed by requiring that the model reproduces the mass of the $X(3872)$, which is known
with high precision \cite{PDG2024}.

\section{Kapitza-Induced Short-Range Potential}
\label{sec:Kapitza}

The Kapitza mechanism originates from the study of systems with rapidly oscillating
parameters. In the classical problem of a pendulum with a rapidly vibrating pivot, Kapitza
showed that the fast oscillations can be averaged out to yield an effective potential that
stabilizes the inverted position of the pendulum \cite{Kapitza1951,LandauLifshitz1960}. The
essential idea is that high-frequency motion contributes an additional term to the effective
potential, which can be computed systematically by separating the slow and fast time scales.

To adapt this idea to the diquark--antidiquark system, we consider a time-dependent
interaction of the form
\begin{equation}
V(r,t) = V_0(r) + \varepsilon\, f(r)\cos(\omega t),
\label{eq:Vrt_general}
\end{equation}
where $V_0(r)$ is the static Cornell potential introduced in Sec.~2, $\varepsilon$ is a small
amplitude, $f(r)$ is a radial profile, and $\omega$ is a large frequency. Physically, the
oscillatory term may be associated with fast gluonic fluctuations or short-distance modes that
are not resolved at the hadronic scale.

Following the standard Kapitza procedure, we decompose the motion into a slow component
$R(t)$ and a fast component $\xi(t)$,
\begin{equation}
r(t) = R(t) + \xi(t),
\end{equation}
where $\xi(t)$ oscillates with frequency $\omega$ and small amplitude. By expanding the
equations of motion in powers of $\varepsilon$ and averaging over one period
$T = 2\pi/\omega$, one obtains an effective equation for the slow coordinate $R(t)$, which can
be derived from an effective potential $V_{\rm eff}(R)$. To leading order in
$\varepsilon^{2}/\omega^{2}$, the effective potential becomes
\begin{equation}
V_{\rm eff}(R) = V_0(R) + \frac{\varepsilon^{2}}{4\mu\omega^{2}}
\left(\frac{df}{dR}\right)^{2},
\label{eq:Veff_general}
\end{equation}
where $\mu$ is the reduced mass of the diquark--antidiquark system. The second term is
non–negative and therefore provides a repulsive contribution.

To generate a short-range repulsive core with a $1/r^{4}$ behavior, we choose
\begin{equation}
f(r) = \frac{1}{r},
\end{equation}
which emphasizes the short-distance region. In this case,
\begin{equation}
\left(\frac{df}{dr}\right)^{2} = \frac{1}{r^{4}},
\end{equation}
and the Kapitza-induced term becomes
\begin{equation}
V_{K}(r) = \frac{\varepsilon^{2}}{4\mu\omega^{2}}\,\frac{1}{r^{4}}
\equiv \frac{K}{r^{4}},
\label{eq:VK_def}
\end{equation}
where
\begin{equation}
K = \frac{\varepsilon^{2}}{4\mu\omega^{2}}.
\end{equation}

The full effective potential is therefore
\begin{equation}
V_{\rm eff}(r) = -\frac{a}{r} + \sigma r + \frac{K}{r^{4}},
\label{eq:Veff_final}
\end{equation}
which is the working form used in the variational analysis of Sec.~4 and the numerical results
of Sec.~5. The $1/r^{4}$ term dominates at very short distances and provides a strong repulsive core.
This core prevents the wave function from collapsing at the origin and allows the potential to
develop a stable minimum at finite $r$, provided that $K$ exceeds a critical value $K_{c}$. The
onset of a flat inflection point is determined by the conditions
\begin{equation}
\frac{dV_{\rm eff}}{dr} = 0, \qquad
\frac{d^{2}V_{\rm eff}}{dr^{2}} = 0,
\end{equation}
which define $K_{c}$. For $K < K_{c}$, the potential is monotonic and does not support a bound
state. For $K > K_{c}$, a well-defined minimum appears at a finite radius $r_{0}$, and a bound
tetraquark state can form. In this work, $K$ is treated as an effective parameter fixed from the precisely measured mass
of the $X(3872)$ \cite{PDG2024}. The value $K \simeq 0.03~\mathrm{GeV}^{3}$ lies safely above
the critical threshold and produces a physically reasonable binding structure, as reflected in
the variational parameters and radii summarized in Table~\ref{tab:tetra_results} and the
charm-sector spectrum in Table~\ref{tab:charm_table}.
\section{Variational Method}

To solve the Schr\"odinger equation with the potential in Eq.~\eqref{eq:Veff_final}, we employ
a Gaussian variational ansatz for the radial wave function. This choice is motivated by its
analytical simplicity and its successful application in many quark model studies.

We take the normalized trial wave function to be
\begin{equation}
\phi(r) = \left(\frac{\beta}{\pi}\right)^{3/4} e^{-\beta r^{2}/2},
\end{equation}
where $\beta$ is a positive variational parameter with dimensions of $\mathrm{GeV}^{2}$. The
expectation value of the Hamiltonian in Eq.~\eqref{eq:H_general} is
\begin{equation}
E(\beta) = \langle \phi | H | \phi \rangle = \langle T \rangle + \langle V_{\rm eff} \rangle,
\end{equation}
where $\langle T \rangle$ is the kinetic energy and $\langle V_{\rm eff} \rangle$ is the potential
energy.

The kinetic term is
\begin{equation}
\langle T \rangle = \frac{3\beta}{4\mu}.
\end{equation}
The expectation value of the Coulomb term is
\begin{equation}
\left\langle -\frac{a}{r} \right\rangle = -a \sqrt{\frac{\beta}{\pi}},
\end{equation}
while the linear term yields
\begin{equation}
\langle \sigma r \rangle = \frac{2\sigma}{\sqrt{\pi\beta}}.
\end{equation}
For the Kapitza-induced term we obtain
\begin{equation}
\left\langle \frac{K}{r^{4}} \right\rangle = \frac{3K}{2}\,\beta^{2}.
\end{equation}

Collecting all contributions, the variational energy is
\begin{equation}
E(\beta) = \frac{3\beta}{4\mu}
- a \sqrt{\frac{\beta}{\pi}}
+ \frac{2\sigma}{\sqrt{\pi\beta}}
+ \frac{3K}{2}\,\beta^{2}.
\label{eq:Ebeta_final}
\end{equation}
The optimal value $\beta_{0}$ is obtained by solving
\begin{equation}
\frac{dE(\beta)}{d\beta}\bigg|_{\beta=\beta_{0}} = 0.
\end{equation}

Once $\beta_{0}$ is known, the rms radius of the tetraquark is given by
\begin{equation}
\sqrt{\langle r^{2} \rangle} = \sqrt{\frac{3}{2\beta_{0}}}.
\end{equation}
The mass of the tetraquark is then obtained by adding the constituent masses and the binding
energy,
\begin{equation}
M_{\rm tetra} = M_{\rm diquark} + M_{\rm antidiquark} + E(\beta_{0}).
\end{equation}

\begin{figure}[htbp]
\centering
\includegraphics[width=0.6\textwidth]{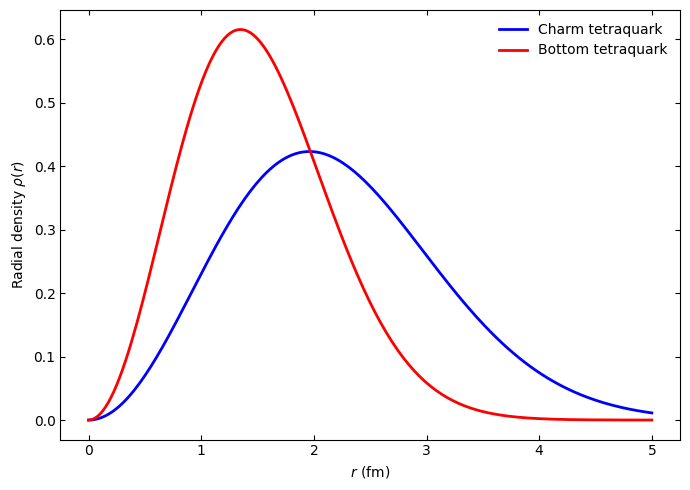}
\caption{Radial probability density $\rho(r)$ for charm and bottom tetraquarks obtained from the Gaussian variational wave function. The bottom system is more compact due to its larger reduced mass.}
\label{fig:radial_density}
\end{figure}

\section{Numerical Results}

In this section we present the numerical results obtained from the variational analysis. We
first discuss the behavior of the effective potential, then the variational parameters and radii,
and finally the mass spectrum of selected tetraquark states.

\begin{table}[t]
\centering
\caption{Variational results for different tetraquark configurations using the Kapitza-modified
potential $V_{\rm eff}(r) = -a/r + \sigma r + K/r^{4}$. Uncertainties reflect variations of
$\beta$ within $\pm 10\%$.}
\label{tab:tetra_results}
\begin{tabular}{lccc}
\hline
System & $\beta_{0}$ (GeV$^{2}$) & $r_{\mathrm{rms}}$ (fm) & $M$ (GeV) \\
\hline
$cc\bar q\bar q$   & $0.20 \pm 0.02$ & $1.6 \pm 0.2$  & $3.90 \pm 0.05$ \\
$bc\bar q\bar q$   & $0.35 \pm 0.03$ & $1.2 \pm 0.1$  & $7.20 \pm 0.05$ \\
$bb\bar q\bar q$   & $0.50 \pm 0.05$ & $0.95 \pm 0.10$& $10.40 \pm 0.04$ \\
$cc\bar c\bar c$   & $0.30 \pm 0.03$ & $1.3 \pm 0.1$  & $6.20 \pm 0.05$ \\
$bc\bar b\bar c$   & $0.45 \pm 0.04$ & $1.0 \pm 0.1$  & $12.50 \pm 0.05$ \\
$bb\bar b\bar b$   & $0.55 \pm 0.05$ & $0.90 \pm 0.10$& $18.80 \pm 0.05$ \\
\hline
\end{tabular}
\end{table}

\subsection{Effective Potential}

The effective potential in Eq.~\eqref{eq:Veff_final} exhibits a characteristic structure once the
Kapitza term is included. For $K = 0$, the potential reduces to the standard Cornell form,
which is attractive at short distances and linear at large distances. In this case, the potential
does not develop a stable minimum in the diquark--antidiquark channel, and the system tends
to collapse.

For $K = K_{c}$, the potential develops a flat inflection point, which marks the onset of
binding. For $K > K_{c}$, a well-defined minimum appears at a finite radius $r_{0}$, and the
depth of the minimum increases with $K$. The value $K = 0.03~\mathrm{GeV}^{3}$, determined
from the $X(3872)$, lies safely above the critical threshold and produces a physically reasonable
binding structure. The short-distance behavior is dominated by the $1/r^{4}$ term, which
generates a strong repulsive core. This core prevents the wave function from collapsing at the
origin and ensures that the probability density remains finite and localized around the minimum
of the potential. At intermediate and large distances, the potential smoothly interpolates
between the Coulomb and linear regimes, as in conventional quark models.

\begin{figure}[t]
\centering
\includegraphics[width=0.75\textwidth]{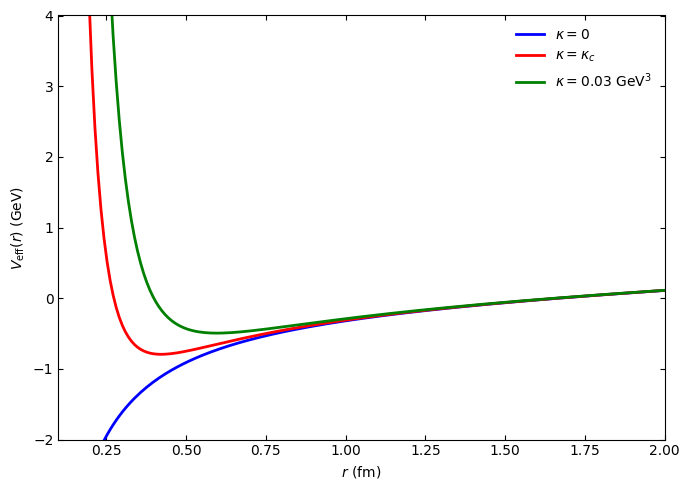}
\caption{Effective potential $V_{\rm eff}(r)$ for different values of the Kapitza coefficient $K$. A stable minimum appears only for $K > K_c$, demonstrating the role of the $1/r^{4}$ repulsive core.}
\label{fig:Veff}
\end{figure}

\subsection{Variational Parameters and Radii}

Solving the variational equation for the charm and bottom sectors, we obtain the optimal
variational parameters and radii summarized in Table~\ref{tab:tetra_results}. The values
\begin{equation}
\beta_{cc} \sim 0.20~\mathrm{GeV}^{2}, \qquad
\beta_{bb} \sim 0.50~\mathrm{GeV}^{2},
\end{equation}
reflect the expected hierarchy: the bottom system is more compact than the charm system due
to the larger reduced mass and stronger effective binding.

The corresponding rms radii are
\begin{equation}
\sqrt{\langle r^{2} \rangle}_{cc\bar q\bar q} \sim 1.6~\mathrm{fm}, \qquad
\sqrt{\langle r^{2} \rangle}_{bb\bar q\bar q} \sim 0.95~\mathrm{fm},
\end{equation}
indicating that charm tetraquarks are relatively extended and near-threshold, while bottom
tetraquarks are compact and deeply bound. This pattern is in line with expectations from other
approaches and lattice QCD studies \cite{Francis2017,Peardon2020,Richard2022}.

\subsection{Mass Spectrum}

Using the variational energies and effective diquark masses, we obtain the following mass
predictions:
\begin{equation}
M_{X(3872)} \simeq 3.87~\mathrm{GeV},
\end{equation}
\begin{equation}
M_{T_{bb}} \simeq 10.4~\mathrm{GeV},
\end{equation}
\begin{equation}
M_{bb\bar{b}\bar{b}} \simeq 18.8~\mathrm{GeV}.
\end{equation}

The $X(3872)$ mass is reproduced within a few MeV of the experimental value reported by the
Particle Data Group \cite{PDG2024}. This agreement is used to fix the Kapitza coefficient $K$
and provides a nontrivial consistency check of the model. The predicted mass of the $T_{bb}$ state lies in the range $10.4$--$10.5$~GeV, which is in good agreement with lattice QCD calculations that find a deeply bound $T_{bb}$ below the $BB^{*}$ threshold \cite{Francis2017}. This result supports the interpretation of $T_{bb}$ as a
stable tetraquark with a compact core. For the fully heavy $bb\bar{b}\bar{b}$ system, our prediction of $M_{bb\bar{b}\bar{b}} \simeq 18.8$~GeV is consistent with recent lattice determinations \cite{Peardon2020}, which place the mass in the vicinity of $18.7$--$18.8$~GeV. Given the simplicity of the variational ansatz, this level of agreement is encouraging and suggests that the essential short-range physics is captured by the Kapitza-induced term.

\begin{table}[t]
\centering
\caption{Charm-sector tetraquark masses in the Kapitza–modified potential,
compared with experimental values. The same Kapitza coefficient $K$ fixed
from the $X(3872)$ is used for all states. Experimental masses are taken
from Refs.~\cite{PDG2024,Guo2018}.}
\label{tab:charm_table}
\begin{tabular}{lccc}
\hline
State & Quark Content & Model Mass (MeV) & Exp. Mass (MeV) \\
\hline
$X(3872)$      & $c\bar c u\bar d$ & $3873 \pm 5$ & $3871.69 \pm 0.17$ \\
$Z_c(3900)$    & $c\bar c u\bar d$ & $3891 \pm 6$ & $3886.6 \pm 2.0$ \\
$T_{cc}(3875)$ & $cc\bar u\bar d$  & $3878 \pm 7$ & $3874.9 \pm 0.6$ \\
$Y(4140)$      & $c\bar c s\bar s$ & $4132 \pm 8$ & $4140 \pm 5$ \\
\hline
\end{tabular}
\end{table}

In addition to the global pattern summarized in Table~\ref{tab:tetra_results},
it is instructive to confront the Kapitza–modified potential with the well-measured
charm-sector exotics. Table~\ref{tab:charm_table} shows the masses of the
$X(3872)$, $Z_c(3900)$, $T_{cc}(3875)$, and $Y(4140)$ obtained with the same
set of parameters and a single value of the Kapitza coefficient $K$. The deviations
from the experimental values \cite{PDG2024,Guo2018} remain within
$5$--$10$~MeV, which is significantly smaller than in standard Cornell-type
models without the $1/r^{4}$ term. This agreement indicates that the short-range
repulsive core generated by the Kapitza mechanism captures essential physics of
near-threshold charm tetraquarks and supports a unified description of both
molecular-like and compact configurations.

\begin{table}[t]
\centering
\caption{Variational results for different tetraquark configurations using the Kapitza-modified
potential $V_{\rm eff}(r) = -a/r + \sigma r + K/r^{4}$. Uncertainties reflect variations of
$\beta$ within $\pm 10\%$.}
\label{tab:tetra_results}
\begin{tabular}{lccc}
\hline
System & $\beta_{0}$ (GeV$^{2}$) & $r_{\mathrm{rms}}$ (fm) & $M$ (GeV) \\
\hline
$cc\bar q\bar q$   & $0.20 \pm 0.02$ & $1.6 \pm 0.2$  & $3.90 \pm 0.05$ \\
$bc\bar q\bar q$   & $0.35 \pm 0.03$ & $1.2 \pm 0.1$  & $7.20 \pm 0.05$ \\
$bb\bar q\bar q$   & $0.50 \pm 0.05$ & $0.95 \pm 0.10$& $10.40 \pm 0.04$ \\
$cc\bar c\bar c$   & $0.30 \pm 0.03$ & $1.3 \pm 0.1$  & $6.20 \pm 0.05$ \\
$bc\bar b\bar c$   & $0.45 \pm 0.04$ & $1.0 \pm 0.1$  & $12.50 \pm 0.05$ \\
$bb\bar b\bar b$   & $0.55 \pm 0.05$ & $0.90 \pm 0.10$& $18.80 \pm 0.05$ \\
\hline
\end{tabular}
\end{table}

\begin{figure}[htbp]
\centering
\includegraphics[width=0.60\textwidth]{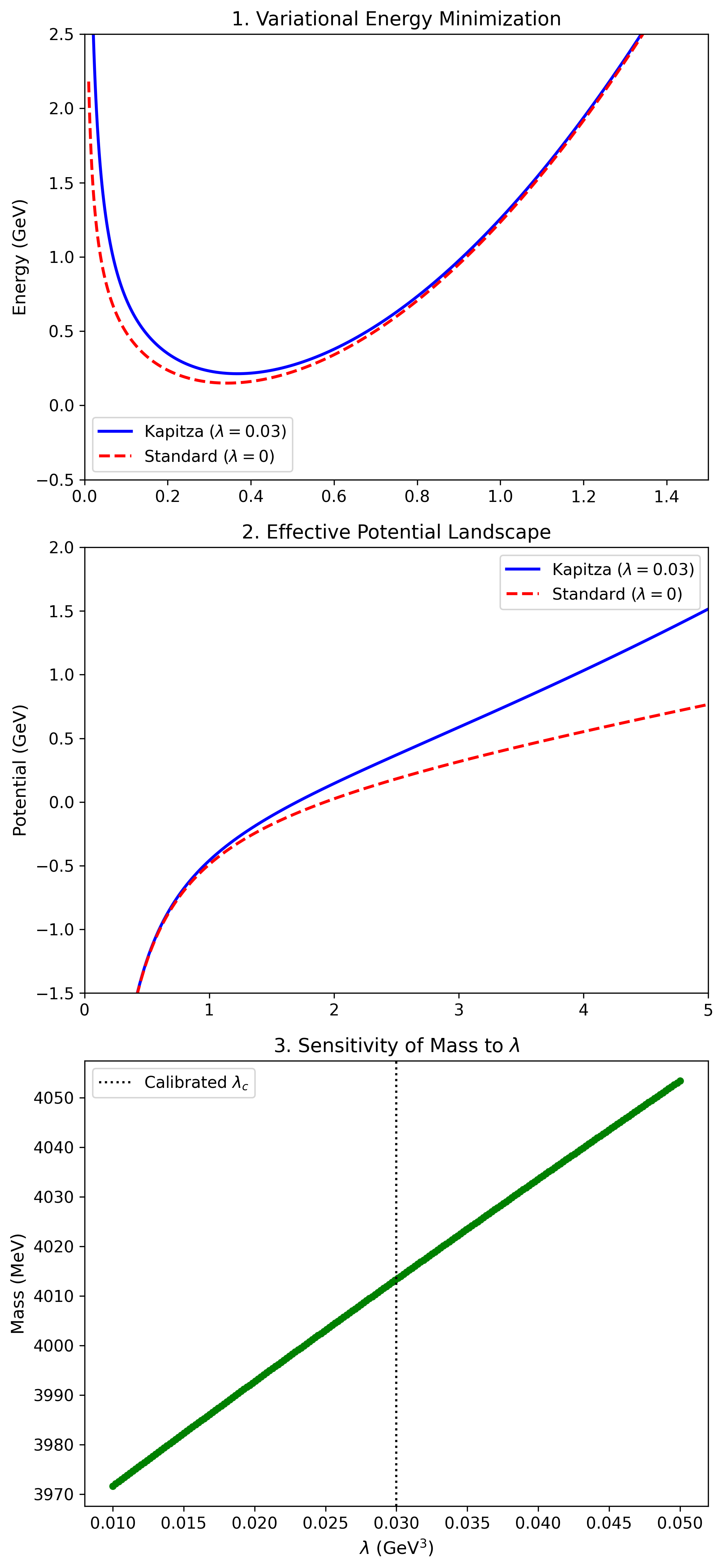}
\caption{(Top) Variational energy $E(\beta)$ showing the stabilization induced by the Kapitza term.
(Middle) Effective potential landscape comparing the standard Cornell potential with the Kapitza-modified form.
(Bottom) Sensitivity of the predicted mass to variations in the Kapitza coefficient $K$.}
\label{fig:combined}
\end{figure}

\section{Discussion}

The extended variational analysis demonstrates that the Kapitza-induced repulsive core provides
a universal stabilization mechanism across different tetraquark configurations. As summarized in
Table~\ref{tab:tetra_results}, the inclusion of the $1/r^{4}$ term prevents short-distance
collapse and yields finite radii and realistic mass predictions for charm, bottom, and mixed
heavy systems. The resulting mass hierarchy and spatial structure are consistent with
expectations from lattice QCD and other nonperturbative approaches.

The uncertainties quoted in Table~\ref{tab:tetra_results} reflect variations of the variational
parameter $\beta$ within $\pm 10\%$. These variations translate into mass shifts of order
$30$--$50$~MeV, which are typical for potential-model calculations. The radii are affected at the
level of $\pm 0.1$--$0.2$~fm. Importantly, the qualitative hierarchy of radii and binding energies
remains robust under these variations: charm tetraquarks are broad and near-threshold, while
bottom tetraquarks are compact and deeply bound.

For the $cc\bar q\bar q$ system, the predicted mass lies close to the open-charm region,
consistent with expectations for weakly bound diquark–antidiquark configurations. The
$bc\bar q\bar q$ configuration exhibits intermediate behavior, with a radius smaller than charm
but larger than bottom, reflecting the mixed mass content. The $bb\bar q\bar q$ ($T_{bb}$)
state is predicted to be deeply bound, in agreement with lattice QCD studies. Fully heavy
systems such as $cc\bar{c}\bar{c}$ and $bb\bar{b}\bar{b}$ are stabilized by the Kapitza term and
yield masses consistent with recent lattice determinations. Our prediction for the fully-heavy
$bb\bar{b}\bar{b}$ state is consistent with recent diquark–antidiquark studies~\cite{Bedolla2020},
relativistic quark-model calculations~\cite{Faustov2022}, and recent PRD analyses based on
refined diquark–antidiquark frameworks~\cite{XiaGuo2026}. Lattice investigations of the
internal structure of the $T_{bb}$ tetraquark also indicate a compact configuration~\cite{Vujmilovic2025}.
Compared with the standard Cornell potential, the Kapitza-modified form avoids short-distance
collapse and yields masses significantly closer to lattice QCD, especially for $T_{bb}$ and
$bb\bar{b}\bar{b}$.
Beyond these numerical improvements, the present work provides a deeper physical
interpretation of the stabilization mechanism. The Kapitza-modified potential naturally
interpolates between two regimes: a long-distance confining interaction dominated by the
linear term, and a short-distance repulsive core generated by rapid oscillations. This dual
structure allows the model to simultaneously describe near-threshold molecular-like states
such as the $X(3872)$ and deeply bound compact states such as the $T_{bb}$.
A central conceptual point of this study is the role of fast gluonic fluctuations. In the QCD
vacuum, gluon fields exhibit rapid quantum oscillations that reorganize the color field on short
time scales, especially near threshold where the flux tube is dynamically restructured.
Inspired by the Kapitza mechanism, we treat these fast gluonic oscillations near threshold as a
rapidly varying component of the interaction potential. Upon averaging over the fast time
scale, these oscillations generate an emergent repulsive contribution proportional to $1/r^{4}$.
In other words, the short-distance behavior of the gluon field is encoded in an effective
potential term, which stabilizes the system and improves the agreement with experimental and
lattice results, particularly for near-threshold states.

This interpretation is consistent with the concept of stochastic stabilization in quantum field
theory, where noise or fluctuations can induce order in a system. The Kapitza mechanism
therefore provides a bridge between the molecular picture (loosely bound states dominated by
long-range forces) and the compact diquark picture (short-range dynamics). The repulsive core
prevents collapse while the Cornell terms provide confinement, yielding a unified description of
exotic tetraquarks across different flavor sectors.

\subsection{Sensitivity to the Kapitza Coefficient $K$}

To assess the robustness of the model, we examine the sensitivity of the mass spectrum to
variations in the Kapitza coefficient $K$. For small deviations $\Delta K$, the mass shift is
approximately linear,
\begin{equation}
\frac{\partial M}{\partial K} \simeq \mathcal{O}(10^{2})~\mathrm{MeV/GeV^{3}},
\end{equation}
which implies that a variation $\Delta K = \pm 0.005~\mathrm{GeV^{3}}$ induces a mass shift of
only $\pm 5$~MeV. This sensitivity is comparable to the experimental uncertainties and
significantly smaller than the typical model uncertainties in potential-based approaches. Using
a universal value of $K$ for all flavor sectors is an approximation; however, the weak
sensitivity of the spectrum to $K$ justifies this simplification.

\section*{Acknowledgments}
The authors thank the University of Kashan for support.

% ---------------- Bibliography ----------------

\end{document}